# A Multi-Stage Deep Learning Framework with PKCP-MixUp Augmentation for Pediatric Liver Tumor Diagnosis Using Multi-Phase Contrast-Enhanced CT

Wanqi Wang[1, 2, #], Chun Yang[1, #], Jianbo Shao[1], Yaokai Zhang[3], Xuehua Peng[1],

Jin Sun[4], Chao Xiong[1, *], Long Lu[2, *], Lianting Hu[1, *]

1. Wuhan Children's Hospital (Wuhan Maternal and Child Healthcare Hospital), Tongji Medical College, Huazhong University of Science and Technology, Wuhan, China
2. School of Information Management, Wuhan University, Wuhan, China
3. School of Electrical Engineering, Wuhan University, Wuhan, China
4. Wuhan Cloud Computing Technology Co., Ltd., Wuhan, China

Pediatric liver tumors are one of the most common solid tumors in pediatrics, with differentiation of benign or malignant status and pathological classification being critical for guiding clinical treatment. While pathological examination is the gold standard for tumor diagnosis, the invasive pathological needle biopsy has notable limitations in children: the highly vascular pediatric liver and fragile tumor tissue raise complication risks such as bleeding and hematoma, especially for small or deep tumors; additionally, young children with poor compliance require sedation or anesthesia for biopsy, increasing medical costs and potential anesthesia-related adverse reactions or psychological trauma. Although numerous efforts have been made to integrate AI into clinical settings, most researchers have overlooked the importance of AI-aided diagnoses of pediatric liver tumors. To establish a non-invasive examination procedure, we developed a multi-stage deep learning (DL) framework for automated pediatric liver tumor diagnosis using multi-phase contrast-enhanced CT.

Two retrospective and prospective cohorts were enrolled. We established a novel PKCP-MixUp data augmentation method to address data scarcity and class imbalance. We also trained a tumor detection model to extract ROIs, and then set a two-stage diagnosis pipeline with three backbones (ViT, ResNet18, DenseNet121) with ROI-masked images. Our tumor detection model has achieved high performance (mAP=0.871, F1=0.866, precision=0.858, recall=0.861), and the first stage classification model between benign tumors and malignant tumors reached an excellent performance (AUC=0.989, accuracy=0.970, precision=0.927, recall=1.000, F1=0.957). Final diagnosis models also exhibited robustness, including benign subtype classification (AUC=0.915, accuracy=0.890, precision=0.388, recall=0.605, F1=0.473) and malignant subtype classification (AUC=0.979, accuracy=0.971, precision=0.950, recall=0.905, F1=0.927). We also conducted multi-level comparative analyses, such as ablation studies on data, augmentation methods, and training pipelines, as well as Shapley-Value-based and CAM-based interpretability analyses of DL models.

This framework fills the pediatric-specific DL diagnostic gap, provides actionable insights for CT phase selection and model design, and paves the way for precise, accessible pediatric liver tumor diagnosis—ultimately optimizing clinical decision-making, improving early intervention, and advancing the paradigm of pediatric radiological AI.

---

# These authors contributed equally.
*Corresponding authors: Chao Xiong (xcpopo11@yeah.net), Long Lu (bioinfo@gmail.com), Lianting Hu (liantinghu@whu.edu.cn)

# Introduction

Pediatric liver tumors account for approximately 1%–2% of all tumors in children[1]. Moreover, nearly two-thirds of pediatric liver tumors are malignant, some instances of which may present with symptoms such as severe abdominal pain and hemorrhagic shock, or even death. Among malignant tumors, hepatoblastoma (HB) and undifferentiated embryonal sarcoma of the liver (UESL) are the most common types[2], while in benign tumors, hemangioma (HH), hepatic mesenchymal hamartoma (HMH), and focal nodular hyperplasia (FNH) are observed[1][3][4]. Treatment strategies vary significantly across these tumor types, making early and accurate diagnosis critical for optimizing therapeutic decisions and improving patient outcomes[5]. Although histopathological biopsy remains the gold standard for diagnosing hepatic masses, it is an invasive procedure that may cause both physical and psychological distress in pediatric patients. It also carries potential risks such as bleeding, infection, and needle tract seeding of tumor cells[4]. Furthermore, the limited sampling range may lead to missed diagnoses.

As a non-invasive modality, computed tomography (CT) plays a vital role in the early screening and diagnosis of liver tumors, both in clinical studies[6][7] and DL-based studies[8][9]. Traditionally, imaging diagnosis relies heavily on radiologists, who must dedicate considerable time and effort to image interpretation. Moreover, the diagnostic process is highly dependent on the clinician's experience and subjective judgment, which may lead to missed or incorrect diagnoses. This poses a challenge for early and accurate detection of pediatric liver tumors and for implementing personalized clinical decision-making. Therefore, there is an urgent need for objective and efficient automated tools capable of accurately distinguishing liver tumors. Such tools will not only alleviate the heavy workload of radiologists but also enable early diagnosis and timely treatment for pediatric patients.

With the rapid advancement of artificial intelligence (AI) technologies, deep learning (DL), as a core subfield, has achieved remarkable progress in medical image analysis and disease diagnosis [4][10][11][12][13][14][15]. DL-based models have been widely applied to imaging-based diagnosis of liver diseases, demonstrating great potential in various tasks such as automatic lesion detection and segmentation, lesion classification, treatment response assessment, and prognosis prediction. Notably, most existing studies have focused predominantly on adult populations, mainly involving common adult liver tumors such as hepatocellular carcinoma, cholangiocarcinoma, and metastatic liver tumors [7[16][17][18][19], while researches on deep learning applications in pediatric liver tumors remain totally scarce, with a limited scope on few subtypes and infancy only[20]. Furthermore, due to substantial differences between pediatric liver tumors and those in adults in terms of disease spectrum and biological characteristics, those DL models for adults cannot be directly implemented in pediatric settings. As a result, the use of deep learning techniques for the automated diagnosis of pediatric liver tumors holds considerable value for future clinical research and applications.

Deep learning models typically require large volumes of high-quality data for effective training. However, in clinical settings for pediatric liver tumors, data collection and annotation are often challenging due to factors such as limited availability of labeled samples and significant class imbalance. These issues pose substantial barriers to achieving high model performance and robustness. Although recent studies have demonstrated the potential of data augmentation techniques to alleviate data scarcity, existing approaches remain limited in their effectiveness. Traditional augmentation methods, such as scaling, rotation, and cropping, provide only marginal improvements in sample diversity[21] and are often unsuitable for tasks that are sensitive to spatial localization[22][23]. On the other hand, generation-model-based synthetic augmentation introduces new challenges, including distributional shifts that deviate from real-world data and increased computational costs for training, as well as additional burdens in manual evaluation and annotation[24][25][26]. Therefore, there is an urgent need to explore more effective and cost-efficient data augmentation strategies in the medical imaging domain.

To address the urgent need for automated and accurate diagnosis of pediatric liver tumors, this study proposes a multi-stage deep learning framework based on multi-phase contrast-enhanced CT imaging. Two real-world cohorts, including both retrospective and prospective ones, were collected and annotated for model training and evaluation. A novel data augmentation method named PKCP-MixUp was introduced to alleviate data scarcity and class imbalance, significantly enhancing classification robustness. Three representative backbones—ViT, ResNet18, and DenseNet121—were systematically compared across three key classification tasks: benign or malignant classification, benign tumor subtype classification, and malignant tumor subtype classification. Moreover, human vs AI comparison, phase ablation, step ablation, CAM-based visualization, and SHAP-based interpretability analysis on the collected cohorts were conducted to enhance models' transparency and explainability. Our study offers a practical and generalizable framework for automated pediatric liver tumor diagnosis, providing novel insights into multi-phase imaging utilization and data-efficient model design.

# Results

### Patient Characteristics

We have collected two cohorts for this study. Cohort A was retrospectively collected from Wuhan Children's Hospital from January 2011 to May 2024. The collection of cohort B was prospectively conducted at Wuhan Children's Hospital between June 2024 and December 2024. The inclusion and exclusion methods are depicted in **Fig. 1**. Patient cohorts comprised a training set, a validation set and a test set. Demographic details, included age and sex distributions, varied across cohorts. Cohort A included 148 patients, indicating an average age of 6.80 years and a female-to-male ratio of 75/69, while cohort B contained 17 patients, indicating an average age of 5.25 years and a female-to-male ratio of 10/7. Additional specifics can be found in **Table 1**.

### Performance of PKCP-MixUp in Tumor Detection

The results shown in **Fig. 2** indicate that the four-phase PKCP-MixUp without class-rebalance configuration achieved the highest overall performance under the threshold of 40% IoU, with mAP (0.871), F1-score (0.866), precision (0.858), and recall (0.861) all exceeding 0.850, demonstrating its strong discriminative capability. Notably, although additional class rebalancing was applied, its performance declined slightly (mAP: 0.829, F1: 0.790, precision: 0.832, recall: 0.754), as shown in **Fig. 2a**. This suggested that over-augmentation or excessive recombination may introduce redundancy, potentially hindering lesion localization. In contrast, as in **Fig. 2b**, performance dropped significantly for the other configurations, indicating that both reduced phase diversity (three-phase PKCP-MixUp: mAP (0.684), F1-score (0.618), precision (0.658), recall (0.563)) and lack of lesion-specific contrast timing (single-phase single-slice: mAP (0.763), F1-score (0.683), precision (0.865), recall (0.619)) negatively impact model accuracy and robustness. These results confirm that leveraging full-phase information in a structured, multi-channel format is crucial for accurate lesion detection, and overly simplified inputs limit the model's ability to capture comprehensive lesion characteristics.

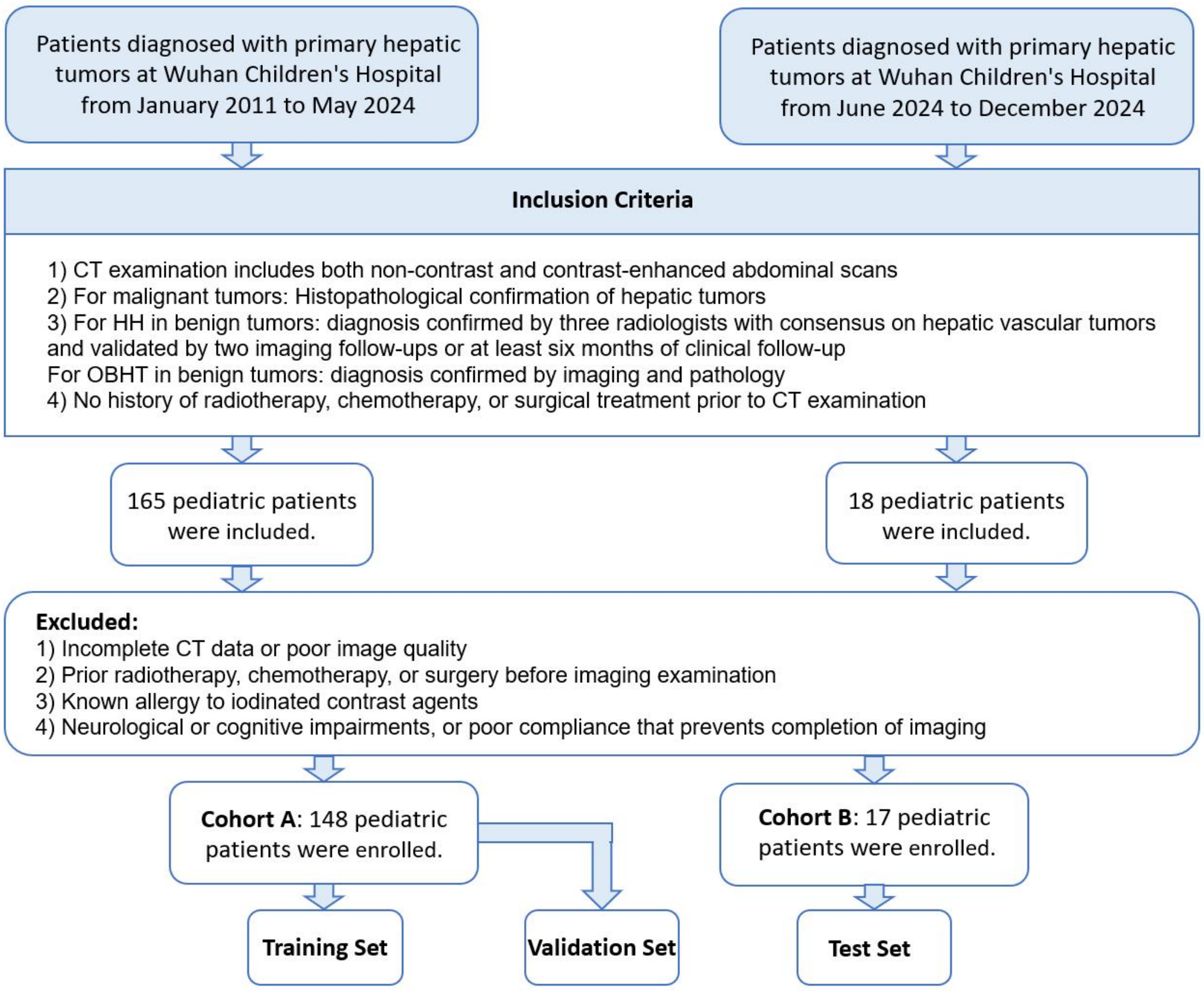

**Fig. 1: The Flowchart of the Cohort Setup.** Two cohorts were obtained from real-world clinical data at Wuhan Children's Hospital.

| Attribute | | Cohort A (p=148, i=1800) | Cohort B (p=17, i=204) |
|---|---|---|---|
| Age | | 6.8, +/-2.8 | 4.1, +/-4.1 |
| Sex (female/male) | | 75/69 | 10/7 |
| Tumor Type | HH | p=77, i=984 | p=14, i=168 |
| | OBHT | p=14, i=168 | p=1, i=12 |
| | HB | p=47, i=600) | p=1, i=12 |
| | OMHT | p=4, i=48 | p=1, i=12 |
| Tumor Size(s) | | 35.771, +/-46.156 | \ |
| Calcification (yes/no) | | 23.00% | \ |
| Bleed(yes/no) | | 0.72% | \ |
| Lung Metastasis(yes/no) | | 1.45% | \ |

**Table 1: Baseline Characteristics of cohorts**, where p and i are separately the numbers of patients and images.

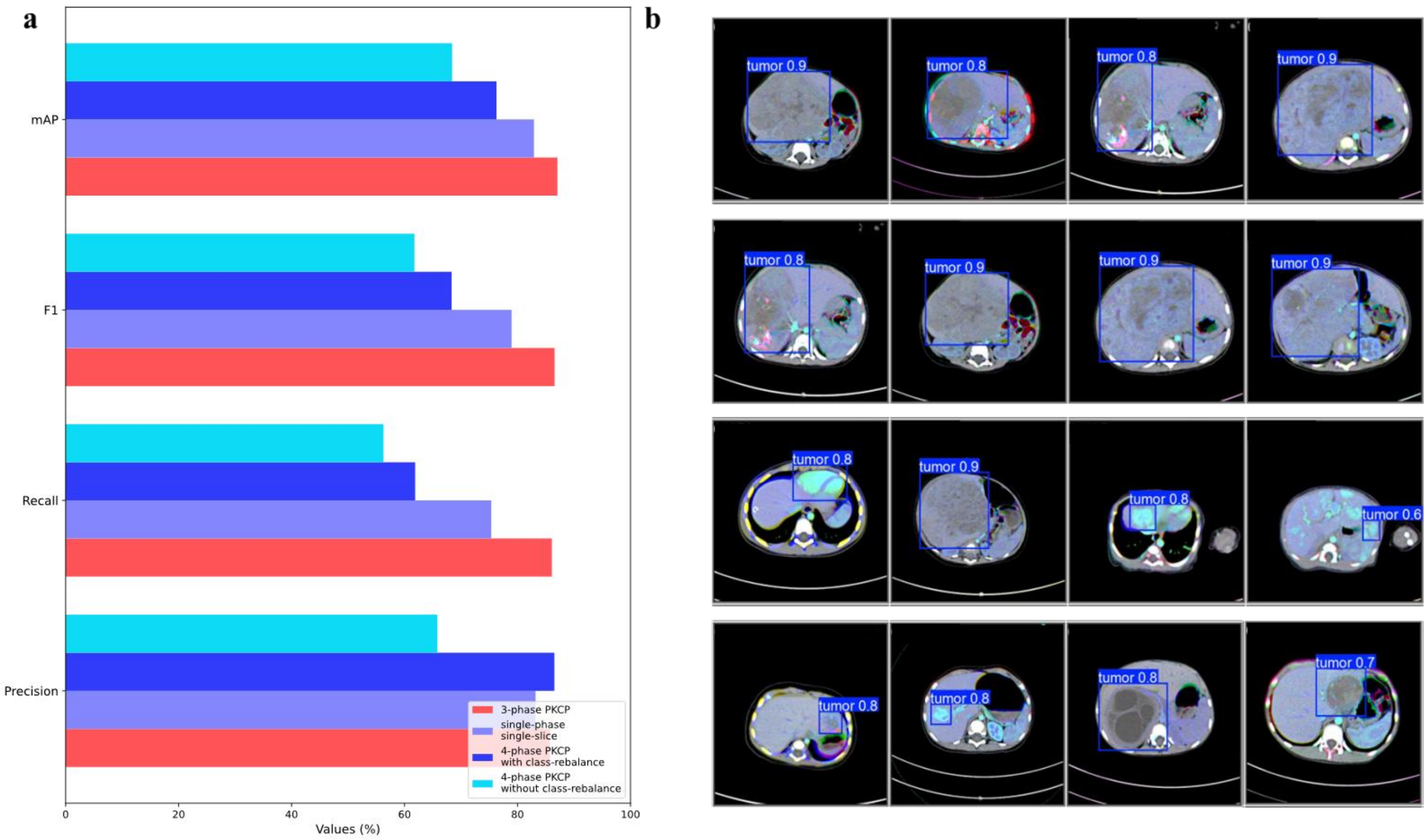

**Fig. 2: The Results of Tumor Detection. a** shows the mAP, F1, recall, and precision results through different configurations. **b** indicates the visualization of YOLOv8's output, the square represents model's prediction of the tumor location, and the number in the label attached to the square represents model's confidence (probability).

## Performance of PKCP-MixUp in Tumor Classification

We trained three backbones to distinguish benign from malignant liver tumors and evaluated the impact of different data augmentation strategies on backbone performance to distinguish malignant tumors from the benign ones. Result details can be found in **Fig. 3a**, **3b**, and **3e**. As **Fig. 3a** shows, across all three backbones, our PKCP-MixUp augmentation method (our aug) consistently achieved superior AUC-ROC results compared to both the traditional augmentation (traditional aug) and the augmentation-free (no aug) settings. Specifically, the ViT model showed the most significant performance boost when trained with our augmentation strategy, achieving an AUC-ROC of 0.989 (95% CI: 0.969-1.000), an accuracy of 0.970 (95% CI: 0.875-1.000), an F1 of 0.957 (95% CI: 0.862-1.000), a precision of 0.927 (95% CI: 0.822-1.000) and a recall of 1.000 (95% CI: 0.906-1.000), compared with traditional augmentation (AUC-ROC: 0.973, 95% CI: 0.943-0.994) and no augmentation (AUC-ROC: 0.958, 95% CI: 0.918-0.985). For ResNet18, our augmentation yielded an AUC-ROC of 0.989 (95% CI: 0.972-0.999), an accuracy of 0.939 (95% CI: 0.845-1.000), an F1 of 0.903 (95% CI: 0.809-0.998), a precision of 0.966 (95% CI: 0.754-0.943) and a recall of 0.848 (95% CI: 0.906-1.000), slightly surmounting the traditional augmentation (AUC-ROC: 0.988, 95% CI: 0.981-0.995). Notably, ResNet18 without augmentation showed a marked drop in AUC-ROC to 0.910 (95% CI: 0.830-0.976), underscoring the importance of effective data expansion in this task. DenseNet121 also benefited from our augmentation strategy and achieved the most comprehensive performance among all three models, with an AUC-ROC of 0.986 (95% CI: 0.965-1.000), an accuracy of 0.980 (95% CI: 0.942-1.000), an F1 of 0.953 (95% CI: 0.965-1.000), a precision of 1.000 (95% CI: 0.963-1.000), and a recall of 0.910 (95% CI:

0.873-0.947), outperforming both comparison settings. It indicated that our PKCP-MixUp augmentation method fitted the case, thus selected as the augmentation method for the subsequent subtype classification tasks.

For the benign tumor subtype classification task shown in **Fig. 3c** and **3f**, DenseNet121 achieved an AUC of 0.915 (95% CI: 0.880-0.950), an accuracy of 0.890 (95% CI: 0.867-0.914), an F1 score of 0.473 (95% CI: 0.449-0.496), a precision of 0.388 (95% CI: 0.365-0.411), and a recall of 0.605 (95% CI: 0.581-0.628). ResNet18 achieved an AUC of 0.911 (95% CI: 0.852-0.965), with a relatively higher recall but significantly lower accuracy, F1 score, and precision compared to DenseNet121. In contrast, ViT performed poorly in this task, with an AUC of only 0.539 (95% CI: 0.453-0.623), indicating limited discriminative capacity. ViT also showed significantly lower F1, precision, and recall, suggesting that it struggled to accurately identify and classify benign liver tumor subtypes in this dataset.

For malignant tumor subtype classification shown in **Fig. 3d** and **3g**, DenseNet121 achieved an AUC of 0.979 (95% CI: 0.933-1.000), an accuracy of 0.971 (95% CI: 0.891-1.000), an F1 score of 0.927 (95% CI: 0.847-1.000), a precision of 0.950 (95% CI: 0.870-1.000), and a recall of 0.905 (95% CI: 0.825-0.985). ResNet18 achieved a slightly higher AUC of 0.988 (95% CI: 0.968-1.000), but with overall performance metrics that were comparatively lower, especially in terms of precision. ViT reached an AUC of 0.876 (95% CI: 0.726-0.994), though its precision and recall remained substantially lower than those of the CNN-based models. These results suggest that while ViT retained some classification ability for malignant subtypes, its performance was less reliable compared to DenseNet121 and ResNet18 in this high-stakes diagnostic task.

**Performance of Different Phases in Tumor Classification**

In clinical practice, liver lesions often exhibit distinct imaging characteristics across different contrast phases, and radiologists typically rely on multi-phase imaging to support diagnostic decisions. Consistent with this approach, our system simultaneously integrates information from multiple CT phases to enhance classification accuracy, providing enhanced support for medical professionals. To evaluate the advantages of incorporating different phases, we conducted ablation experiments. The results can be found in **Fig. 4**.

To assess the contribution of each phase to diagnostic performance, ablation experiments have been conducted on all three classification tasks: benign or malignant diagnosis, benign subclassification, and malignant subclassification. As illustrated in **Fig. 4d**, the four-phase combination consistently demonstrated the highest diagnostic performance across all tasks, with ROC-AUCs of 0.97, 0.91, and 0.98, respectively. Notably, PC provided maximal diagnostic value in benign or malignant diagnosis, with a drop of 0.57 in ROC-AUC from its absence, followed by AP (a drop of 0.04). For benign tumor subclassification and malignant tumor subclassification, PVP presented the most irreplaceable role with a drop of 0.24 in the former and 0.18 in the latter.

**Performance of Different Steps in Tumor Classification**

In comparison with our two-step approach, we conducted a one-step experiment for further illustration. As illustrated in **Fig. 5a**, the model yielded a macro-average AUC of 0.762 and a weighted-average AUC of 0.837 with evident class imbalance. Among the four classes, OMHT showed the highest diagnostic performance, with an AUC of 0.888 (95% CI: 0.816-0.938). HH also achieved a relatively high AUC of 0.805 (95% CI: 0.590-0.908). In contrast, the classification performance for OBHT and HB was lower, with AUCs of 0.718 (95% CI: 0.388-0.916) and 0.638 (95% CI: 0.056-0.918), respectively. Collectively speaking, the one-step approach lacked comprehensiveness across all categories, compared with a two-step approach.

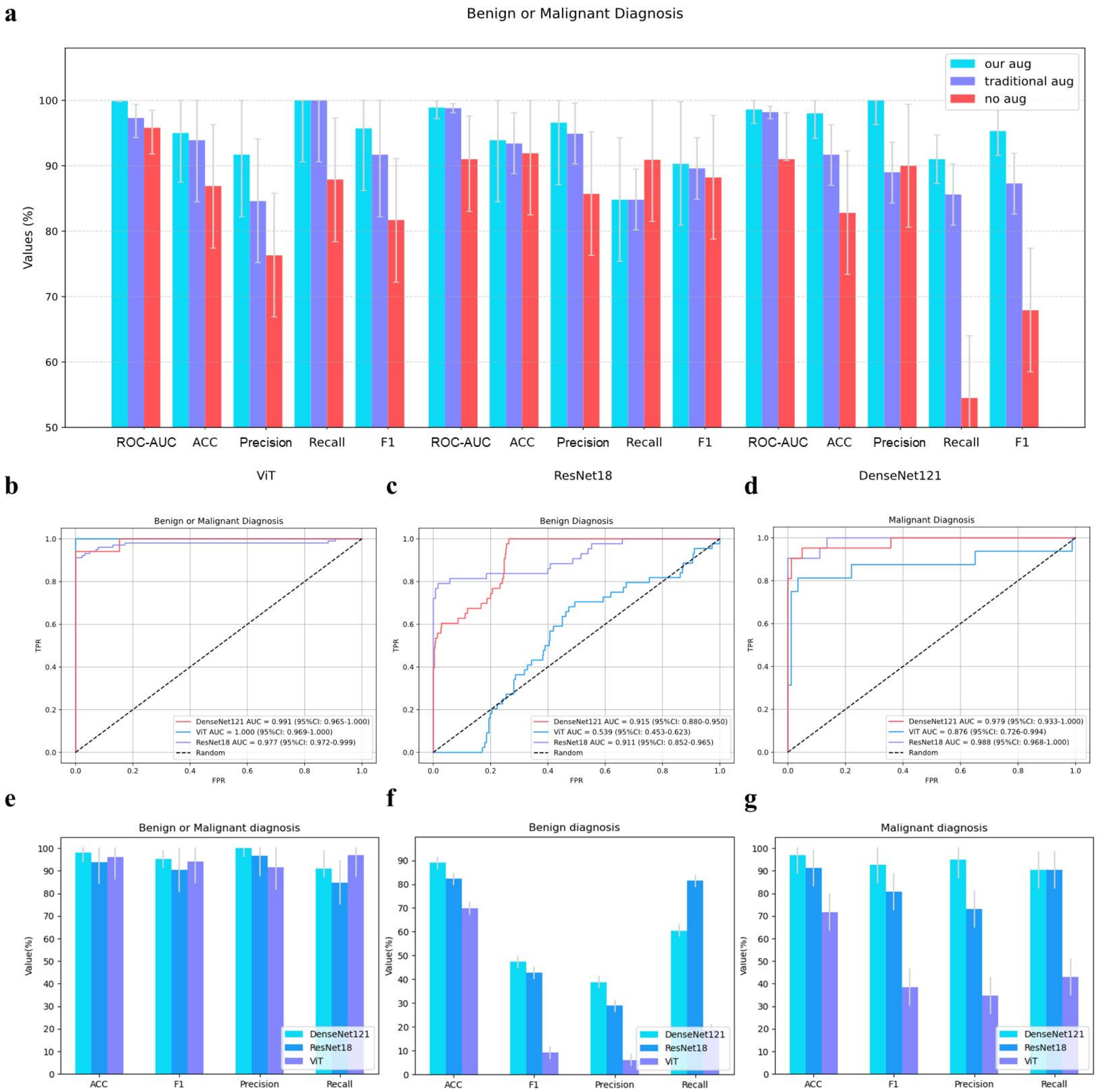

**Fig. 3: Classification Model Performance on Validation Set. a** shows ROC-AUC, accuracy (ACC), Precision, Recall and F1 values in the benign or malignant diagnosis based on different augmentation methods. **b-d** respectively show the ROC curves of benign or malignant classification, benign subtype classification, malignant subtype classification, while **e-g** represent the corresponding performance metrics of them.

## Performance of Clinical Experience in Tumor Classification

To further investigate the potential in radiologists' experience in AI-assisted clinical diagnoses, we trained two more models based on CatBoost and Multilayer Perceptron (MLP), integrating visual features extracted from previous vision models and structured information from radiology reports. As illustrated in **Fig. 5j**, the vision-only DenseNet121 model achieved the highest AUC (0.980) in the benign or malignant classification task, substantially outperforming the multimodal CatBoost (AUC: 0.820) and multimodal MLP (AUC: 0.810) models. However, in the subtype classification tasks, multimodal models exhibited superior performance. Multimodal CatBoost achieved near-perfect AUCs of 1.000 for both benign and malignant subtype classification, exceeding the performance of the vision-only model, while multimodal MLP only surmounted the vision model on the benign subtype task with an AUC of 0.990. These findings indicate that while vision-only models are highly effective in binary discrimination, the integration of manual radiology features from radiologists' experience through multimodal models provides substantial benefit for finer-grained subtype classification.

Then we calculated the Shapley values of multimodal features on multimodal CatBoost for subtype tasks to find out the detailed contributions of them. As shown in **Fig. 5k**, age and density were listed as the top 10 features in the benign diagnosis, indicating that younger patients or patients with a higher level of uniformity in density were more likely to be diagnosed as HH. Unlike the benign diagnosis, in **Fig. 5l**, the malignant diagnosis found the importance rankings of structured information are as follows: density, area and age. Lower level of uniformity in density, smaller area and lower age contributed to the model's diagnosis of HB.

**Comparison with Radiologists**

We used the validation set to compare the diagnostic ability of the best model in overall classification (DenseNet121) with that of two junior radiologists, each with less than three years of experience in radiological diagnosis. DenseNet121 achieved an AUC of 0.915, with a sensitivity of 0.605, specificity of 0.711, and an overall accuracy of 0.890. In contrast, junior radiologist 1 achieved a sensitivity of 0.786, specificity of 0.500, and accuracy of 0.737. Radiologist 2 had a sensitivity of 1.000, a specificity of 0.000, and the same accuracy of 0.737. These results demonstrate that the diagnostic performance of DenseNet121 was substantially superior to that of both junior radiologists. No statistically significant difference was observed between the diagnostic performances of the two radiologists. The corresponding ROC curves are presented in **Fig. 5b**.

Similarly, for the classification of malignant hepatic tumor subtypes, DenseNet121 achieved an AUC of 0.979, with a sensitivity of 0.905 (95% CI: 0.825–0.985), a specificity of 0.889, and an accuracy of 0.971 (95% CI: 0.891–1.000). In comparison, Radiologist 1 achieved a sensitivity of 0.833, a specificity of 0.000, and an accuracy of 0.625, while Radiologist 2 obtained a sensitivity of 1.000, a specificity of 0.167, and an accuracy of 0.375. Again, DenseNet121 demonstrated significantly higher diagnostic performance than the junior radiologists. No statistically significant difference in diagnostic accuracy was observed between the two radiologists. The corresponding ROC curves for malignant tumor subtype classification are shown in **Fig. 5c**.

**Prospective Dataset Evaluation**

On the test set for benign-versus-malignant classification, DenseNet121 achieved an accuracy of 0.978 (95% CI: 0.902-1.000), an F1 score of 0.976 (95% CI: 0.912-1.000), a precision of 0.977 (95% CI: 0.901-1.000), and a recall of 0.976 (95% CI: 0.924-1.000). ResNet18 also performed well, with all four metrics consistently above 0.950. ViT achieved slightly lower results across all metrics but still maintained accuracy, F1 score, precision, and recall above 0.930. Both CNN-based models demonstrated stable and high classification performance for this task.

On the test set for benign subtype classification, DenseNet121 achieved an accuracy of 0.941 (95% CI: 0.850-1.000), an F1 score of 0.667 (95% CI: 0.575-0.758), a precision of 1.000 (95% CI: 0.909-1.000), and a recall of 0.500 (95% CI: 0.409-0.591). ResNet18 demonstrated a comparable set of scores. However, ViT exhibited reduced performance across all metrics, particularly in precision (0.444, 95% CI: 0.353-0.536) and recall (0.667, 95% CI: 0.575-0.758). Overall, DenseNet121 outperformed the other models in prospective benign-malignant differentiation, maintaining high precision while achieving acceptable recall.

On the test set for malignant subtype classification, DenseNet121 achieved excellent scores of 100% across the overall performance metrics, while ResNet18 and the ViT model performed much more poorly, suggesting that they struggled to maintain discriminative capability for malignant subtypes in the test set. To sum up, DenseNet121 consistently exhibited the best generalization and sensitivity in this critical diagnostic task.

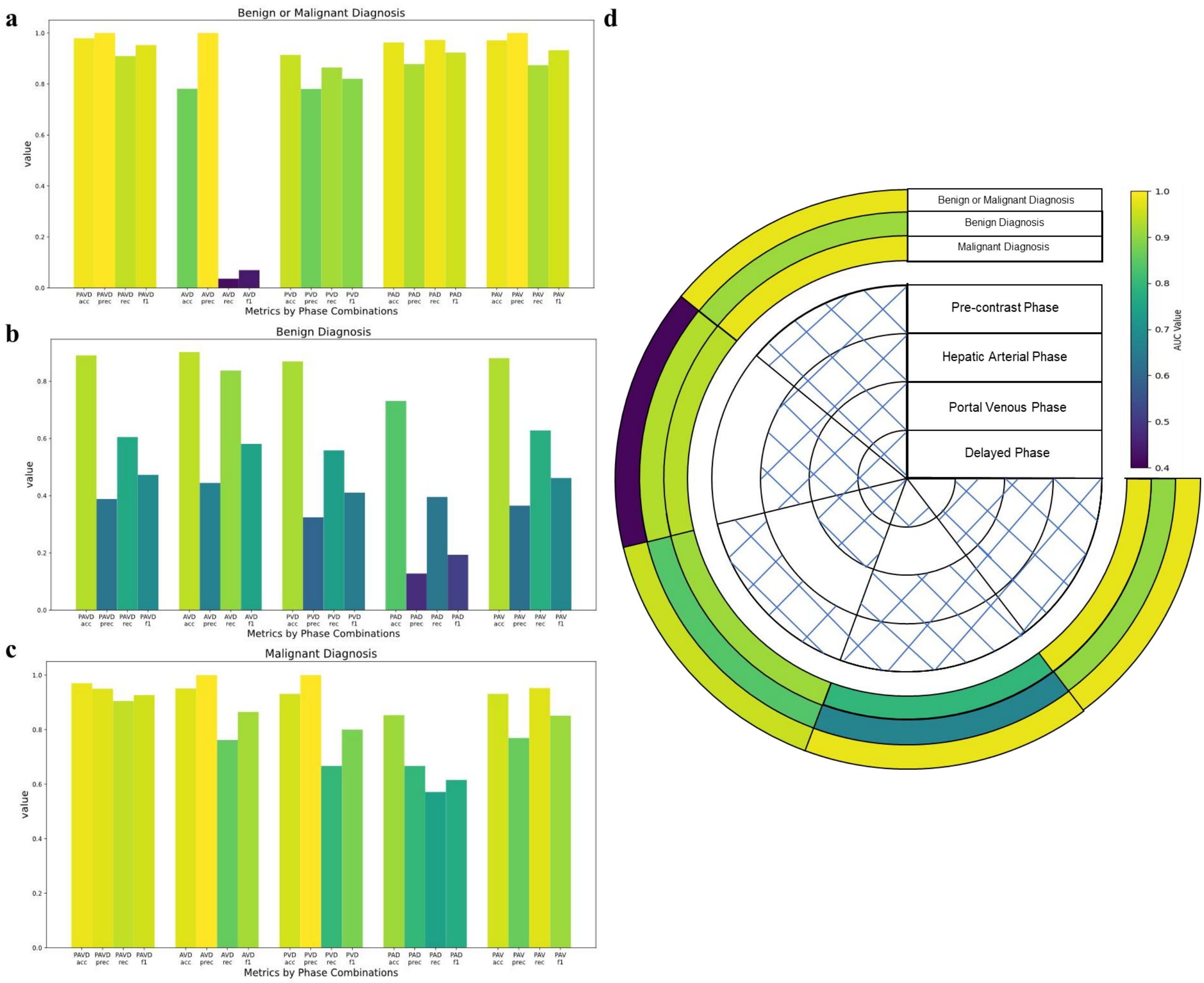

**Fig. 4: Results of Multi-Phase Ablation Study. a-c** represent the model's performance metrics on the validation set with different phase combinations. Among them, P is short for pre-contrast phase, A is short for hepatic arterial phase, V represents portal venous phase, and D is the delayed phase. **d** indicates the model's ROC-AUC across 3 classification tasks. The outer annulus shows the ROC-AUC scores in alignment with the color legend on the right, while the inner annulus shows the usage of phases (if grid-lined, the corresponding phase is used).

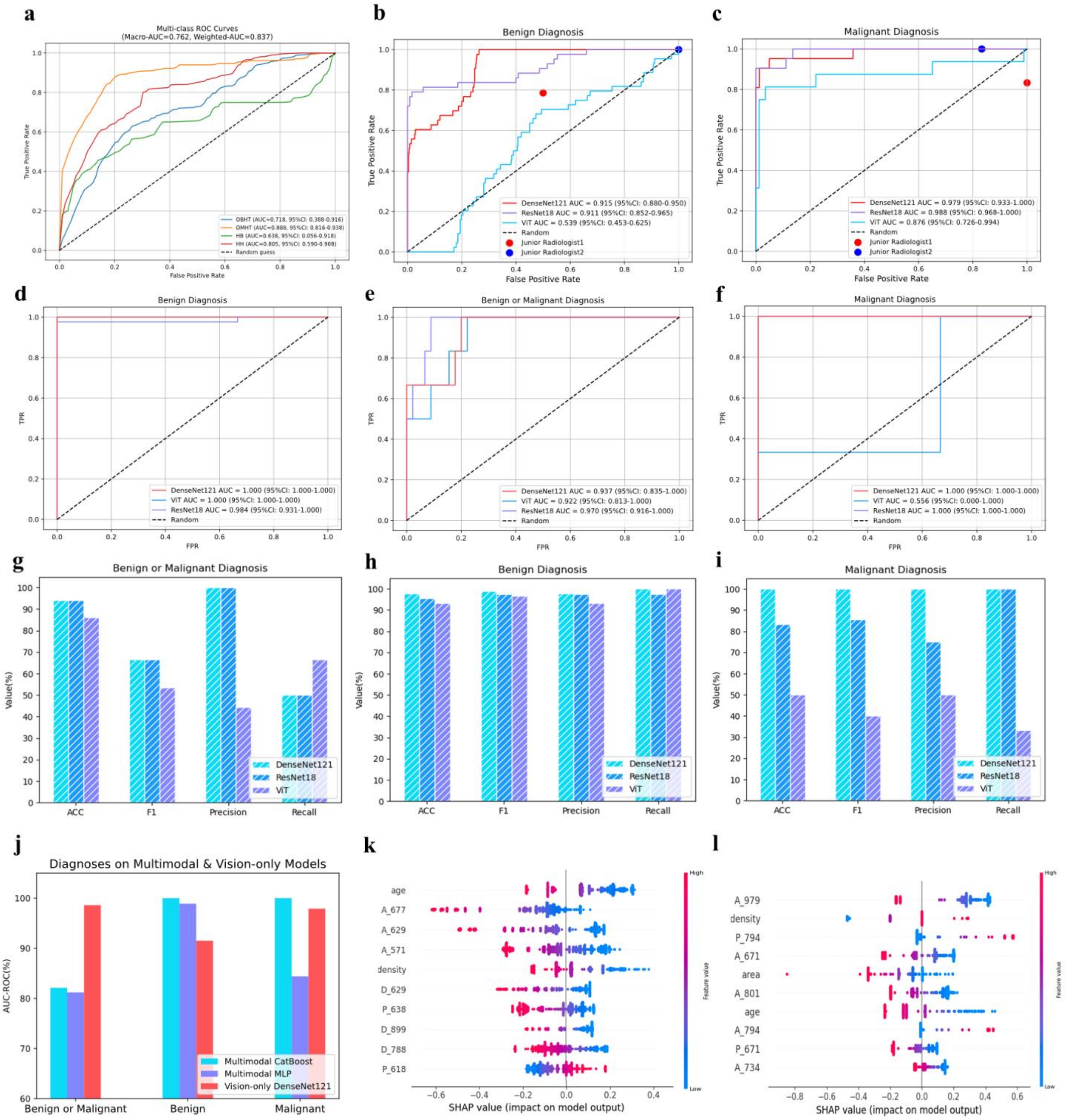

**Fig. 5: Performance of Different Steps in Tumor Classification, Comparison with Radiologists, Prospective Dataset Evaluation, and Performance of Clinical Experience in Tumor Classification. a** shows the overall ROC curves of the one-step approach. **b-c** show the ROC curves of models in two subtype classification tasks in comparison with those of the junior radiologists (denoted as dots). **d-l** illustrate the models' performance on Cohort B, the prospective dataset. **j** indicates ROC-AUC values on the validation set based on different model and modality configurations. **k-l** show the Shapley values calculated for the best multimodal model of each subtype task. A higher ranking suggests the feature is more significant in contributing to the result. Each dot represents a sample case in the study, while different colours represent relative values and the x-axis shows whether it has a negative or positive impact on the output probability of the binary classification. Image features extracted from vision models are named with a structure-phase in short, plus the sequence of overall features in the corresponding image.

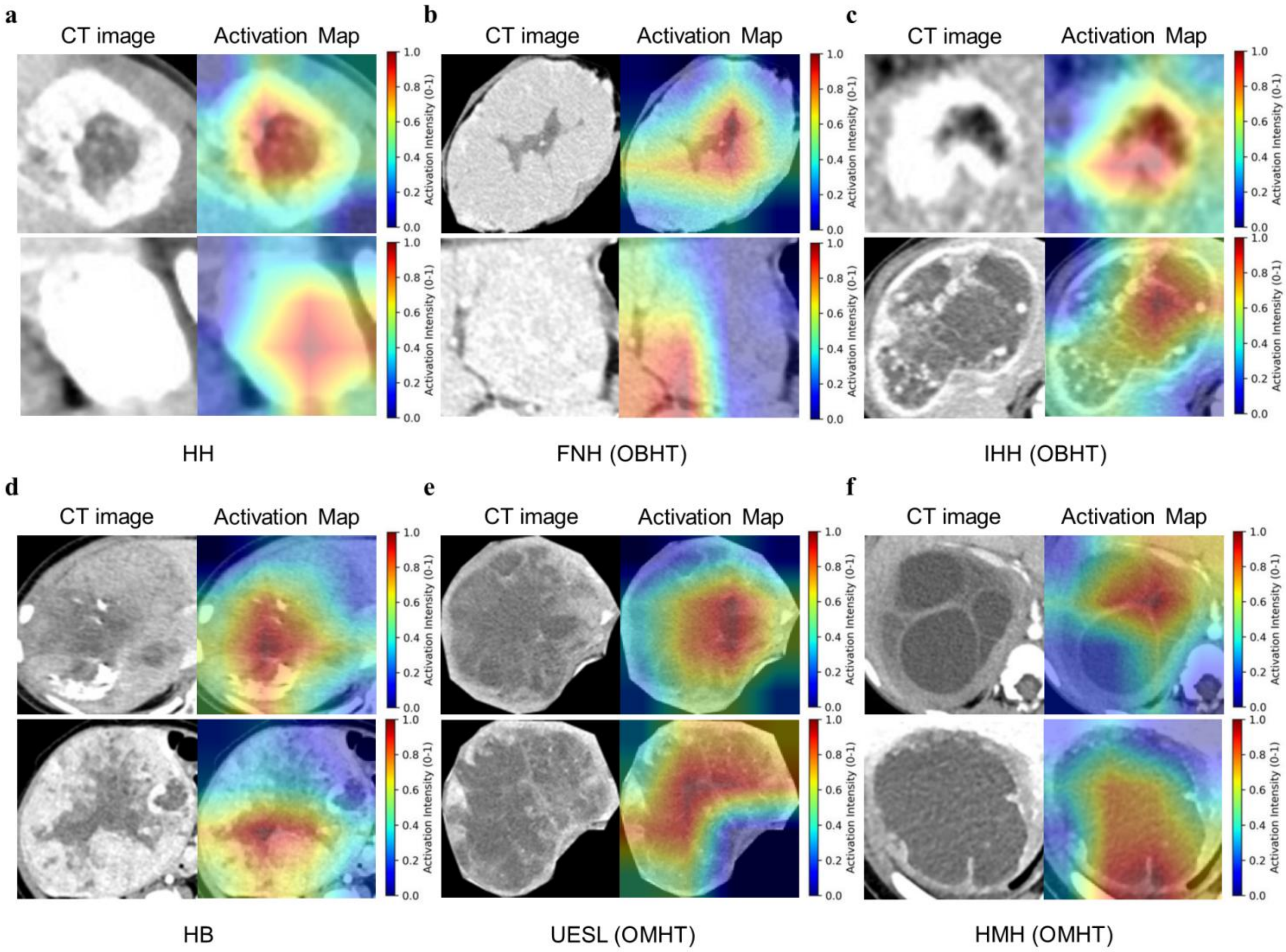

**Fig. 6: The CT images and the corresponding class activation maps of each type of liver tumors in this study.** Each tumor type is presented with two patients' CT images. Left images are original CT images (in phase AP), while on their right show corresponding heatmaps, the value of which is correlated with the model's area attention.

## CAM (Class Activation Maps)-based Deep Learning Analysis

To better explain the deep learning model, we conducted an analysis by professional radiologists on activation maps. Class activation maps (CAMs) are generated by computing the activation level of each pixel in the image by the model, revealing the areas of focus within the image. **Fig. 6** shows that the model pays more attention to lesion areas relative to normal liver tissue to distinguish between different subtypes.

The visualization of hepatic hemangioma (HH) lesions is shown in **Fig. 6a**. The left column presents original images from different cases of HH lesions in dynamic contrast-enhanced CT scans, while the right column displays the corresponding liver tumor regions identified by the deep learning model, with the lesion areas predominantly highlighted in red, indicating the model's focused attention. The visualization of focal nodular hyperplasia (FNH) lesions is shown in **Fig. 6b**. On the left are the original CT images of FNH lesions, where the characteristic central stellate scar is visible; on the right are the corresponding results from the deep learning model, in which the red-highlighted area is concentrated around the central scar. The visualization of infantile hepatic hemangioma (IHH) lesions is shown in **Fig. 6c**. The left column presents original CT images from different IHH cases, and the right column shows the liver tumor regions detected by the model. The visualization of hepatic mesenchymal hamartoma (HMH) lesions is shown in **Fig. 6f**. The left image displays the original CT scan of an HMH lesion, characterized by a multilocular cystic structure with thin cyst walls. During contrast-enhanced imaging, enhancement of the cyst wall and internal septa is visible. The right image shows the model's identification of the tumor lesion, with red regions concentrated on the enhanced cyst walls and septa. The visualization of hepatoblastoma (HB) lesions is shown in **Fig. 6d**. The left image presents the original CT scan of an HB lesion, where heterogeneous enhancement is observed. The right image shows the tumor areas identified by the model, with red regions corresponding to areas of high vascular density and metabolic activity. The visualization of undifferentiated embryonal sarcoma lesions (UESL) is shown in **Fig. 6e**, with the left image showing the original CT scan and the right image displaying the corresponding tumor region identified by the deep learning model.

# Disussion

As shown in **Fig. 3b** and **Fig. 3e**, in the benign or malignant classification task for liver tumors, ViT demonstrated excellent performance in the benign or malignant classification task for liver tumors. Its high recall makes it particularly advantageous for early detection of malignant tumors, helping to reduce false negatives. For suspected liver tumor patients, using ViT for an initial benign or malignant determination, followed by further tumor-type confirmation, could improve screening efficiency and optimize medical resource allocation, offering a novel approach for objective imaging-based diagnosis of pediatric liver tumor malignancy. As shown in **Fig. 3c-3d** and **Fig. 3f-3g**, in the benign tumor subtype classification task and the malignant tumor subtype classification task, DenseNet121 outperformed ViT and ResNet18 in the validation and test sets, demonstrating excellent ability to differentiate between HH and OBHT. DenseNet121 showed strong robustness against dataset variability and sufficient learning abilities in the face of relatively small datasets, enhancing its diagnostic accuracy in clinical settings. Given its outstanding performance on both validation and test cohorts, DenseNet121 was selected as the optimal model for both benign and malignant tumor subtype classification tasks.

In the benign tumor subtype classification task, DenseNet121 significantly outperformed two junior physicians on accuracy, suggesting strong potential for clinical application in screening and diagnosing pediatric HH. The model's diagnostic performance in the malignant tumor subtype classification task, with an accuracy of 0.971, significantly exceeded that of the two junior physicians. Through deep analysis and feature extraction from CT image data, the model accurately identified HB and effectively distinguished it from OMHT. HB is the most common malignant liver tumor in childhood; early diagnosis and prompt intervention are critical for improving prognosis[27][28]. The application of DenseNet121 opens new opportunities for achieving this goal, potentially offering novel clinical pathways for HB management and bringing renewed hope to affected children and their families.

In routine image interpretation, radiologists typically review the pre-contrast phase (PC) and three contrast-enhanced phases—AP, PVP and DP—to arrive at a final imaging diagnosis. This study simulated the conventional radiology workflow using four-phase CT image combinations. By sequentially excluding one phase, we generated four combinations: AVD (AP, PVP, DP), PVD (PC, PVP, DP), PAD (PC, AP, DP), and PAV (PC, AP, PVP), and input these into the optimal model for each classification task to compare diagnostic performance across different combinations.

For the benign or malignant classification task (shown in **Fig. 4a** and **Fig. 4d**), the full four-phase combination PAVD achieved the highest diagnostic performance among all combinations. It enabled accurate differentiation between benign and malignant liver tumors, and this result is consistent with the findings by Asaka et al.[29]. Multiphase contrast-enhanced imaging reveals tumor characteristics across different scanning phases. The AP reflects the vascular supply of tumors, which helps the model recognize the distinct enhancement patterns between malignant tumors and benign tumors. The DP shows differences in blood supply between the two types of tumors: malignant tumors clear contrast rapidly, leading to reduced enhancement, while benign tumors retain contrast, resulting in high attenuation. The PC does not have enhancement characteristics, but it provides critical supplementary information such as tumor morphology, density differences, location, and size—all of which help distinguish benign and malignant tumors.

Moreover, for the benign tumor subtype classification task shown in **Fig. 4b** and **Fig. 4d**, the three-phase combination AVD outperformed the full four-phase combination PAVD. It improved the differentiation between HH and OBHT. This result suggests that the PC may interfere with the model's judgment. The reason is that common pediatric benign tumors—including HH, FNH, and HMH—show isodensity or hypodensity relative to normal liver parenchyma in the pre-contrast phase. These subtle density differences can confuse the model, which further underscores the value of enhanced CT for accurate differentiation of pediatric benign liver tumors.

As for the malignant tumor subtype classification task shown in **Fig. 4c** and **Fig. 4d**, the three-phase combination PAV outperformed other phase combinations. All other combinations that included the DP showed varying degrees of decreased diagnostic performance. This indicates that DP may hinder the model's ability to distinguish between HB and OMHT. This observation aligns with the enhancement characteristics of tumors: HB exhibits heterogeneous arterial enhancement and rapid contrast washout in PVP and DP, and lesions appear isodense or hypodense relative to normal liver parenchyma in DP. Similarly, HCC, one kind of OMHT, shows "rapid in and out" enhancement with rapid contrast washout in DP, making lesion enhancement less conspicuous compared to the surrounding parenchyma. Therefore, focusing on the PAV imaging combination significantly enhances diagnostic accuracy when distinguishing between different types of malignant liver tumors.

As illustrated in **Fig. 4l**, within the malignant subtype classification task, age acts as a critical factor in the diagnostic process of the CatBoost model: the younger the patient, the more likely the model is to classify the case as HB. This result aligns with clinical observations—HB is predominantly diagnosed in patients under 4 years of age, whereas the onset ages of the other malignant tumor subtypes are relatively later. Specifically, the incidence of HCC is highest in patients older than 10 years, while that of UESL peaks in the 5–10 years age group.

In summary, although the full four-phase PAVD combination offers strong performance in benign or malignant classification tasks, certain three-phase combinations may emphasize key distinguishing features, enabling deep learning models to achieve superior classification performance[30]. In clinical practice, deep learning models can assist radiologists in making preliminary judgments about tumor malignancy and then, based on diagnostic needs and actual conditions, focus attention on specific phases to further refine tumor subtype diagnosis and improve accuracy.

Through CAM analysis, we observed that the deep learning model accurately focused on the pathological core features of different pediatric liver tumors, with activation patterns highly consistent with their histological and imaging characteristics. As a highly vascular lesion[31], HH typically shows progressive centripetal contrast enhancement on dynamic CT and homogeneous enhancement in the delayed phase. The model's CAM exhibited widespread, uniform activation across the entire lesion (highlighted uniformly), indicating it learned HH's key feature of rich, evenly distributed vascularity, as shown in **Fig. 6a**. Pathologically, FNH is characterized by abnormal hepatocyte arrangement and a central fibrous stellate scar with radiating septa and malformed feeding arteries[32][33][34][35]. The CAM showed high activation specifically in the central stellate scar region, demonstrating the model's ability to capture this distinctive pathological marker of FNH, as shown in **Fig. 6b**. HB, the most common malignant pediatric liver tumor[36], has tumor cells arranged in clusters or trabeculae, uneven vascular supply (denser in the center), and heterogeneous CT enhancement with necrosis or hemorrhage[37]. The CAM in **Fig. 6d** displayed concentrated high activation in the tumor center (corresponding to vascular-rich, metabolically active areas) and low activation in the periphery—reflecting the model's learning of HB's core features: central hypervascularity and heterogeneous enhancement. In contrast, HMH usually present as predominantly solid lesions or cystic-solid/multilocular cystic masses with fibrous septa that enhance on contrast CT[38]. The model's CAM showed high activation in the solid components (septa and stroma) and minimal activation in cystic areas. This confirms the model learned HMH's key feature of enhanced solid septa. Classic cases can be found in **Fig. 6f**.

This study has several limitations. First, it is a single-center study, lacking validation using large-scale, multi-center datasets. As a result, the model's performance has not been evaluated across diverse populations from different geographic regions or healthcare settings, which limits the generalizability of the findings. Second, the number of enrolled cases in this study was relatively small. Future studies should include larger sample sizes to further validate the robustness and generalizability of the deep learning models. Third, there was a class imbalance in the tumor types included in this study. Hepatic hemangioma (HH) accounted for most benign tumors, while hepatoblastoma (HB) dominated the malignant tumor group. This imbalance limits the model's ability to perform precise subtype classification for both benign and malignant tumors. Fourth, the study only included primary liver tumors, excluding secondary hepatic malignancies in pediatric patients. This restricts the model's applicability in more complex and diverse clinical scenarios. Fifth, the current study focused solely on imaging data, without incorporating general clinical information such as laboratory findings or tumor markers. Future work should emphasize the integration of multi-modal data from larger datasets and investigate optimal model designs in order to achieve more accurate recognition and classification of pediatric liver tumors.

# Methods

## Data Collection

We have collected two cohorts for this study. Cohort A was retrospectively collected from Wuhan Children's Hospital from January 2011 to May 2024. The collection of cohort B was prospectively conducted at Wuhan Children's Hospital between June 2024 and December 2024. Patients were included if they met all of the following conditions: (1) underwent both pre-contrast and multiphase contrast-enhanced abdominal CT examinations; (2) had a diagnosis of a primary hepatic tumor confirmed by histopathology or long-term imaging follow-up (≥6 months for benign tumors); (3) had no history of other malignant tumors or concurrent extrahepatic malignancies; (4) had no prior treatment, including surgery, radiotherapy, or chemotherapy, before imaging. Patients were excluded based on the following conditions: (1) CT scans were incomplete or of poor quality; (2) Presence of severe comorbidities (e.g., congenital malformations, metabolic diseases) that could interfere with imaging interpretation; (3) Mental or physical conditions that rendered imaging unfeasible or incomplete.

All CT scans were performed using Siemens multi-detector CT systems (SOMATOM Force and SOMATOM Definition AS) at Wuhan Children's Hospital. Patients were scanned in the supine position, head-out and feet-first, with imaging performed from the diaphragm to the inferior margin of the liver. The protocol included a pre-contrast phase (PC) followed by three contrast-enhanced phases: hepatic arterial phase (AP), portal venous phase (PVP), and delayed phase (DP). For uncooperative young children, 10% chloral hydrate (0.25-0.5 mL/kg) was administered orally or rectally. Scanning parameters included 120 kV tube voltage, 270 mAs tube current, 1.0 mm slice thickness, and CARE Dose 4D modulation. A nonionic contrast agent (iopamidol, GE Pharmaceutical) was injected via the elbow vein using a Bayer injector. Arterial, portal venous, and delayed phases were acquired at 10-25 s, 50-60 s, and 2 min after injection, respectively.

CT scans were stored in DICOM format and reviewed using the CARESTREAM PACS system. For each radiology test of a patient, 3 representative slices centered on the primary lesion were selected from each CT phase by experienced radiologists, resulting in a total of 12 (3 slices × 4 phases) slices for one radiology test of a patient. Using the image export function of the hospital's PACS (Picture Archiving and Communication System), thin-slice images from all four phases were exported in JPEG format for convenient model training. The window settings were 40 HU window level and 250 HU window width for the non-contrast phase and 45 HU window level and 290 HU window width for contrast-enhanced phases. Phase alignment was achieved via synchronization, with manual correction applied if needed. The exported slices all went through an anonymization process to ensure patients' privacy.

To improve diagnostic accuracy, original CT slices were preprocessed to isolate regions of interest (ROIs) by minimizing irrelevant background. Initially, ROIs were first manually annotated by a junior radiologist with LabelMe. The annotations were then confirmed by a senior radiologist. If there was a conflict, the two radiologists would discuss till reaching a consensus.

Corresponding imaging reports contain the following structured information: age; sex, male or female; calcification; bleed; lung metastasis; tumor size, width*heights; density, labeled by 5 hirarchical levels; lesion location, categorized into 7 labels (Multiple Intrahepatic Lesions, Left Hepatic Lobe, Right Hepatic Lobe, Quadrate Lobe of the Liver, Caudate lobe of the liver, Hepatic Hilum, Translobar); enhancement features with 15 kinds of labels (details are illustrated in Appendix Table 1). Among these, the first five were directly extracted from patients' digital radiology reports, while the rest were measured and restructured by one junior radiologist and confirmed by one senior radiologist. The structured information was then encoded into features.

Liver tumors were classified into four categories: (1) hepatic hemangiomas (HH) in benign hepatic tumors; (2) other benign hepatic tumors (OBHT), including infantile hepatic hemangioma (IHH), focal nodular hyperplasia (FNH) and hepatic mesenchymal hamartoma (HMH); (3) hepatoblastomas (HB) in malignant hepatic tumors; (4) other malignant hepatic tumors (OMHT), including undifferentiated embryonal sarcoma of the liver (UESL) and Hepatocellular Carcinoma (HCC). This classification scheme was adopted to reflect both the clinical prevalence and the pathological nature of pediatric liver tumors. The detailed classification criteria of liver tumors are as follows: (1) Histopathological confirmation based on surgical specimens or biopsy was used for OBHT, HB, and OMHT[22][39]. (2) Imaging-based criteria were applied to HH, characterized by peripheral nodular enhancement during the arterial phase with progressive centripetal fill-in on the portal venous and delayed phases, or by homogeneous, marked enhancement during the arterial phase (comparable to aortic enhancement), followed by high attenuation on delayed images[40]. It was also additionally confirmed either by consensus diagnosis among three radiologists or by consistent imaging findings across at least two modalities during a minimum six-month follow-up period.

Cohort A was randomly split to a training set and a validation set by a ratio of 3/7, while cohort B was used as the test set. The training set included a total of 104 patients with 1248 CT slices and the validation set included a total of 44 patients with 528 CT scans. Our test set contained 17 patients with 204 CT slices. The split of Cohort A adhered to Patients' ID, which was unique for each patient, to prevent data overlap in the training set and validation set.

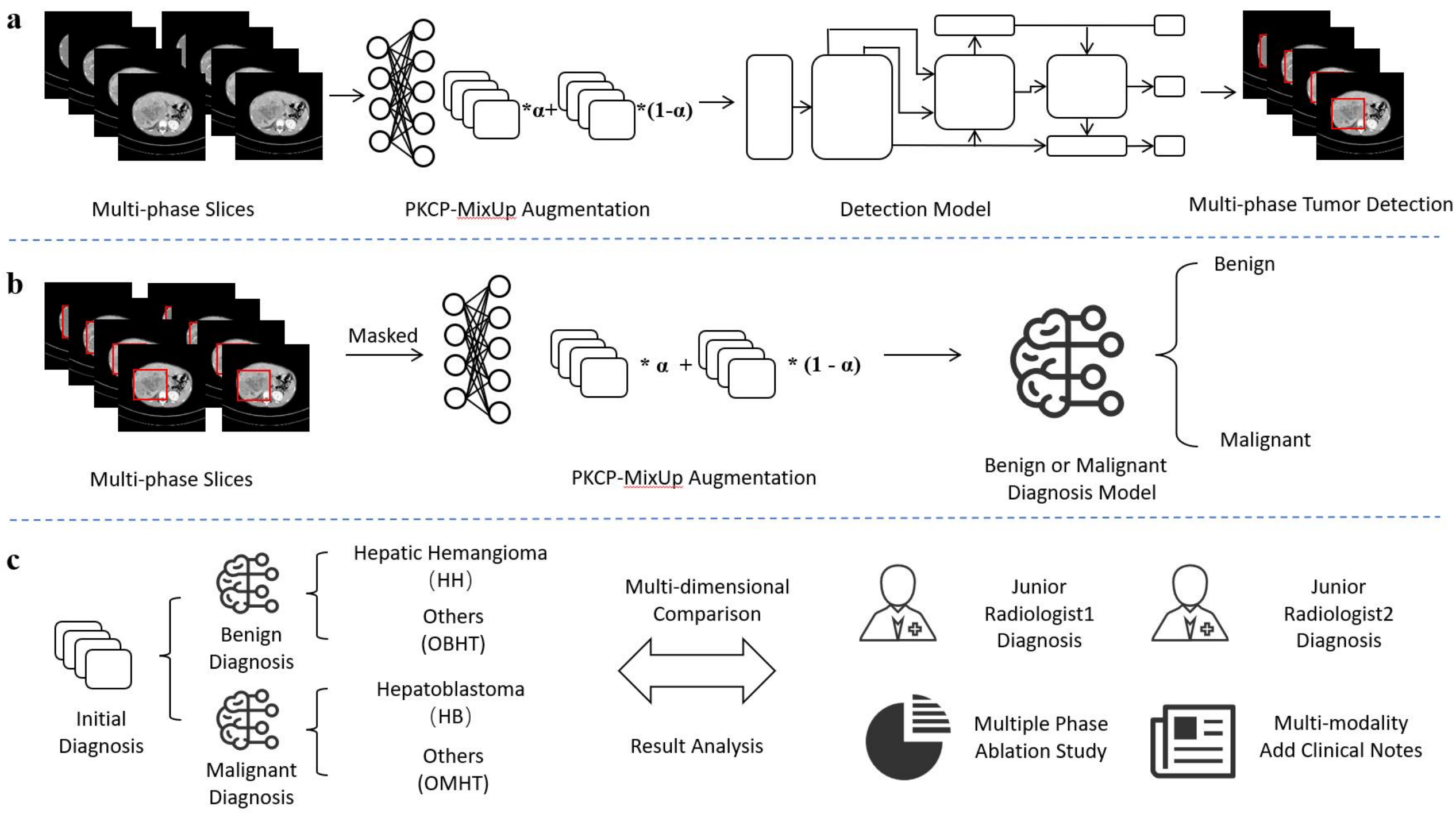

**Fig. 7: The Pipeline of the Study.** This study follows trained a pediatric liver tumor detection YOLOv8 model

before training three 2-step diagnosis models with our proposed data augmentation method (PKCP+MixUp). Model performance were tested on two sets with radiologists. Further model explainability was validated on a multi-phase ablation study, CAM-based heatmaps, and SHAP-based multi-modal models with hand-crafted features.

**Data augmentation**

To mitigate the risk of overfitting due to limited training samples and the class imbalance, we implemented a novel data augmentation named PKCP-MixUp by combining exquisite augmentation technique MixUp with another new method named Phase-wise K-slice Cartesian Product (PKCP), as shown in **Fig. 7a** and **Fig. 7b**.

**1) PKCP Augmentation**

We have initially selected 12 slices containing visible lesions in the data collection process. These slices were categorized by imaging phase into four subgroups corresponding to the pre-contrast phase (PC), hepatic arterial phase (AP), portal venous phase (PVP), and Delayed Phase (DP), with each phase containing three slices. The mathematical illustrations are as follows. Denote the CT slices of one radiology case as a set of phase-specific slices:

$$\boldsymbol{S} = \{s_{p,k} | p \in \{1,\dots,P\},\ k \in \{1,\dots,K\}\} = \begin{pmatrix} s_{1,1} & \cdots & s_{1,k} & \cdots & s_{1,K} \\ \vdots & & & & \vdots \\ s_{p,1} & & s_{p,k} & & s_{p,K} \\ \vdots & & & & \vdots \\ s_{P,1} & \cdots & s_{P,k} & \cdots & s_{P,K} \end{pmatrix} \tag{1}$$

where $P$ represents the number of CT phases (e.g., PC, AP, PVP, DP); $K$ represent the number of available lesion slices in each phase; $s_{p,k}$ represents the $k$-th lesion slice in $p$th phase.

A single composite image set $c$ requires one slice from each phase and can be represented by the following formulation:

$$c = \left(s_{1,k^{(1)}}, s_{2,k^{(2)}}, \dots, s_{P,k^{(P)}}\right)^{\mathrm{T}}, k^{(1)}, k^{(2)}, \dots, k^{(P)} \in \{1, 2, \dots, K\}. \tag{2}$$

From each case, we can obtain a total of $N$ composite image sets, denoted as $\boldsymbol{C}$ :

$$\boldsymbol{C} = (c_1, c_2, \dots, c_N). \tag{3}$$

First, we addressed the limited availability and the class imbalance through class-based sample expansion. For each radiological study involving OBHT or OMHT lesions, intra-patient slice recombination for minority classes. To generate synthetic composite samples, we randomly selected one slice from each phase and recombined them to form a four-phase composite image set. Each image set for these minority classes is denoted as $c_i$ :

$$c_i = \left(s_{1,j^{(1)}}, s_{2,j^{(2)}}, \dots, s_{P,j^{(P)}}\right)^{\mathrm{T}} \tag{4}$$

where $i \in \{1, 2, \dots, N\},,\ j^{(1)}, j^{(2)}, \dots, j^{(P)} \in \{1, 2, \dots, K\}$.

The total number of possible composite image sets generated from one case is:

$$N_{\text{minor}} = K^P. \tag{5}$$

Given that $K = 3$ and $P = 4$, 81 distinct multiphase composite image sets were constructed per patient case in the OBHT and OMHT groups, greatly enhancing the representation of these undersampled categories.

In contrast, for the majority classes (HH and HB), the 12 lesion slices were partitioned into three depth-consistent groups (e.g., upper, middle, and lower third of the lesion volume). Each group contained one slice from each of the four phases, and all slices in a given group shared the same slice depth. Each image set for these majority classes is denoted as $c_i$ :

$$c_i = \left(s_{1,j^{(1)}}, s_{2,j^{(2)}}, \ldots, s_{P,j^{(P)}}\right)^{\mathrm{T}} \tag{6}$$

where $i \in \{1, 2, \ldots, N\}$, $j^{(1)} = j^{(2)} = \ldots = j^{(P)}$, $j^{(1)} \in \{1, 2, \ldots, K\}$.

The total number of possible composite image sets generated from one case is:

$$N_{major} = K. \tag{7}$$

Thus, with $K = 3$, 3 distinct composite image sets were constructed per patient case in the HH and HB groups, enhancing the representation of these categories while restraining their sample size from the minority classes.

Each synthetic image set was subsequently fused into a multichannel image, where each channel corresponded to a different phase (PC, AP, PVP, PD). To further enable our method for the tumor detection task, we computed the minimum bounding box that enclosed all lesion areas present in the constituent slices for the associated lesion annotation. The mathematical illustrations are as follows:

$$A_{\text{composite}} = \min\left\{A_i \supseteq \bigcup_{k=1}^{4} A_k\right\} \tag{8}$$

where $A_k$ represents the lesion annotation in phase k.

**2) MixUp augmentation**

Aside from the approach to expand the sample size of the minority, we further applied MixUp augmentation[41] to enhance data diversity in the feature space. This technique involved pixel-level interpolation between two randomly selected images using a mixing coefficient α, which was sampled from a Beta distribution, Beta(α=2, β=2). This process introduced nonlinear variation in the training samples, effectively improving generalization by smoothing decision boundaries across classes. Specific mechanisms can be described as follows:

$$\hat{c} = \lambda c_I + (1-\lambda)c_C, \quad \hat{y} = \lambda y_I + (1-\lambda)y_C \tag{9}$$

where $(c_I, y_I) \in (\mathcal{D}, \mathcal{S_J})$, $(c_C, y_C) \in (\mathcal{D}, \mathcal{S_C})$. $\mathcal{D}$ refers to the training dataset, $\mathcal{S_J}$ and $\mathcal{S_C}$ refer to instance-based and class-based sampling respectively. $c_I$ and $c_C$ separately correspond to one composite image set as illustrated above in PKCP. $\hat{c}$ refers to the final input image and $\hat{y}$ refers to the final input label.

In this study, we have augmented the training set with both methods above. The validation set also employed our method, given the limited data. Correspondingly, to ensure models' diagnoses would land on one specific radiology test in the validation set, the multiple digit outputs of newly recombined multichannel images would be averaged if they originated from the same radiology report ID. Nevertheless, the test set did not employ either data augmentation approach.

**3) Experimental Configurations**

For clear illustration of the effectiveness of our novel data augmentation technique, other types of data configurations were designed as contrast groups: (1) single-phase single-slice images; (2) three-phase PKCP images, in which corresponding lesion slices were from randomly selected three phases; (3) four-phase PKCP without class-rebalance, where all classes were deemed as the majority classes through PKCP process; (4) four-phase PKCP images without any augmentation methods; (5) four-phase PKCP images with only traditional augmentation methods such as random rotation, flipping, etc.; (6) four-phase PKCP images with our PKCP-MixUp method.

**Tumor Detection**

The performance of deep learning models in tumor classification relies heavily on their ability to capture informative features from imaging data. Excessive background content can obscure lesion regions and reduce the model's focus on relevant structures. So a YOLOv8[42]-based model using annotated images was trained for automated ROI extraction, as shown in **Fig. 7a**. Compared to earlier YOLO versions, YOLOv8 introduces several architectural improvements, including an anchor-free detection head and a revised loss function structure. These enhancements contribute to improved object localization accuracy and reduced computational overhead.

YOLOv8 employs an anchor-free detection paradigm, allowing the model to directly predict bounding box centers and sizes without relying on pre-defined anchor boxes. This is especially beneficial for lesions with high inter-patient variability in size and shape.

The model also integrates C2f (Cross-Stage Partial Fusion) modules for more efficient feature aggregation and deeper receptive fields with fewer parameters. The mechanism of efficiency enhancement can be represented by the following mathematical equation:

$$Y = \text{Concat}\Big(X_1, \mathcal{B}_n\big(\mathcal{B}_{n-1}(\dots\mathcal{B}_1(X_2)\dots)\big)\Big), \tag{10}$$

$$X \in R^{C\times H\times W},\ X = [X_1, X_2],\ Y \in R^{C\times H\times W}$$

where $X_1$ is a subset of channels passed directly, $X_2$ represents the remaining channels processed through $n$ Bottleneck blocks $\mathcal{B}$. $Concat$ indicates channel-wise concatenation. $\mathcal{B}_i$ are residual bottleneck layers with shared parameters or repeated structure.

The object detection task is optimized using a composite loss function, formally expressed as:

$$\mathcal{L}_{total} = \lambda_{box} \cdot \mathcal{L}_{box} + \lambda_{obj} \cdot \mathcal{L}_{obj} + \lambda_{cls} \cdot \mathcal{L}_{cls} \tag{11}$$

where $\mathcal{L}_{box}$ is based on Complete-IoU (CIoU) loss for precise bounding box regression, $\mathcal{L}_{obj}$ is a binary cross-entropy loss indicating whether an object exists, $\mathcal{L}_{cls}$ is a multi-class cross-entropy loss over predefined tumor classes, and λ terms are weighting factors empirically determined.

Given the small size and complex anatomical context of pediatric liver lesions, YOLOv8's refined localization capability and low false-positive rate make it a highly effective backbone for automated lesion detection in this study.

In the training settings, lesion annotations were all labeled as class 0: "tumor". Images were resized to 1280*1280. Other configurations included input channels = 4, mixup = 0.6, epochs = 300, early-stopping = True, with a default learning rate.

**Tumor Classification**

For the classification task, we adopted a two-stage methodology to ensure the best accuracy of the diagnostic output. First, we trained a model to classify the malignant tumors (HB & OMHT) from the benign ones (HH & OBHT). Then, two other models were developed to classify detailed categories in malignant tumors and benign tumors separately. The pipeline is shown in **Fig. 7b**.

To explore the effect of different paradigms on classification performance, we evaluated three representative backbone architectures: DenseNet121, ResNet18, and Vision Transformer (ViT).

DenseNet's is a densely connected convolutional neural network that encourages feature reuse and efficient gradient flow. Each layer receives the feature maps of all preceding layers, which can be expressed as:

$$x_l = H_l([x_0, x_1, \dots, x_{l-1}]) \tag{12}$$

where $x_l$ is the output feature map of the $l$-th layer, $[x_0, x_1, \dots, x_{l-1}]$ is concatenation of feature maps from all preceding layers, $H_l(\cdot)$ is composite function of batch normalization, ReLU activation, and convolution at layer $l$.

ResNet18 offers residual connections, making it robust to vanishing gradients, allowing reliable training on limited CT data without overfitting. The output of each residual block is formulated as:

$$x_l = F(x_{l-1}) + x_{l-1} \tag{13}$$

where $x_{l-1}$ represents the input feature map to the residual block, $F(\cdot)$ represents residual mapping function, $x_l$ represents the output feature map after residual addition.

Vision Transformer (ViT) leverages global self-attention mechanisms to capture long-range spatial dependencies. The standard ViT forward pass for one attention layer is expressed as:

$$\text{Attention}(Q, K, V) = \text{softmax}\left(\frac{QK^{\top}}{\sqrt{d_k}}\right)V \tag{14}$$

where $\mathrm{Q, K, V}$ represent Query, Key, and Value matrices, obtained by linear projection of the input patch embeddings, $d_k$ is the dimensions of the Key vectors, $\text{softmax}(\cdot)$ is the function applied along the key dimension to compute attention weights, $\text{Attention}(\mathrm{Q, K, V})$ weighted sum of value vectors, forming the output of the attention layer.

The irrelevant parts of the images were masked before the images were sent to models, which were adapted to 4-channel image inputs. Each model was trained on the same preprocessed datasets described earlier and evaluated under identical conditions for fair comparison. We initialized the network parameters by loading pretrained network layer parameters from the ImageNet dataset. The network was trained using random gradient descent and cross-entropy loss for weight adjustment and algorithm optimization. The initial learning rate was set to 0.001. To mitigate overfitting, batch normalization and a weight decay rate of 0.0001 were implemented during training. A batch size of 128 and a rectified linear unit (ReLU) activation function were used. All codes were implemented using Python 3.10.15. The packages or software comprised Pytorch 2.1.0 for model training and testing, CUDA 12.1for GPU acceleration.

**Explainability Analysis**

As shown in **Fig. 7c**, to validate the effectiveness of the proposed model and enhance its interpretability, we conducted a series of comparative and analytical experiments encompassing the following components: (1) Comparison between single-stage and two-stage classification frameworks to assess the benefit of hierarchical tumor categorization; (2) Performance comparison between the model and junior radiologists, evaluating diagnostic concordance and potential clinical applicability; (3) Ablation study based on different phase images to investigate the contribution of multi-phase information to model performance; (4) Heatmap-based attention visualization with Grad-CAM to explore the spatial focus of the model within tumor regions; (5) SHAP (SHapley Additive exPlanations) -based interpretability analysis, incorporating radiologist-derived features into the model to examine the role of clinical experience in diagnostic reasoning.

**Statistical Analysis**

We evaluated the classification model using statistical measures such as the accuracy, sensitivity, specificity, precision, recall, F1-score (F1), area under the curve (AUC), and receiver operating characteristic (ROC) curve. We employed a default threshold of 0.5, a widely accepted standard for binary classification problems. For multiclassification tasks, we chose the category with the highest probability as the prediction based on the softmax classifier output. By analyzing the accuracy, AUC, F1, recall, and precision, we gained insights into the model's ability to accurately classify instances, handle imbalanced datasets, and strike a balance between true positives, false positives, true negatives, and false negatives, depending on the specific requirements of the application. For each metric, the point estimate was first determined. The standard error of the mean was computed as the error metric. Using a 95% confidence level and degrees of freedom equal to sample size minus 1, the lower and upper bounds of the confidence interval were derived from the t-distribution, with the point estimate as the mean and the standard error as the scale parameter. The resulting 95% CI was presented as the interval between these two bounds.

# Data Availability

The data that support the findings of this study are available from the corresponding author upon reasonable request.

# References

[1] Nasher O, Woodley H, Alizai S, et al. Hepatic benign and malignant masses in children: a single UK tertiary centre experience[J]. Pediatr Surg Int, 2022,38(12):2019-2022.

[2] Chung E M, Lattin G E, Cube R, et al. From the Archives of the AFIP: Pediatric Liver Masses: Radiologic-Pathologic Correlation Part 2. Malignant Tumors[J]. RadioGraphics,2011,31(2):483-507.

[3] Zeng Guang, Tian Zhiyao, Yang Chun, et al. Primary Liver Tumors in Children: Key Imaging Diagnostic Points and Research Progress [J]. Chinese Journal of Radiology. , 2023,57(8):936-940. (in Chinese)

[4] Challagundla Y, Tunuguntla T S C, Tunuguntla S G, et al. Convolutional neural network-based classifiers for liver tumor detection using computed tomography scans[J]. INNOVATIONS IN SYSTEMS AND SOFTWARE ENGINEERING, 2023:7.

[5] Koh K, Namgoong J, Yoon H M, et al. Recent improvement in survival outcomes and reappraisal of prognostic factors in hepatoblastoma[J]. Cancer Med, 2021,10(10):3261-3273.

[6] Honda H, Matsuura Y, Onitsuka H, et al. Differential diagnosis of hepatic tumors (hepatoma, hemangioma, and metastasis) with CT: value of two-phase incremental imaging[J]. AJR. American journal of roentgenology, 1992, 159(4): 735-740.

[7] Oliva M R, Saini S. Liver cancer imaging: role of CT, MRI, US and PET[J]. Cancer imaging, 2004, 4(Spec No A): S42.

[8] Zhang H, Luo K, Deng R, et al. Deep learning-based CT imaging for the diagnosis of liver tumor[J]. Computational Intelligence and Neuroscience, 2022, 2022(1): 3045370.

[9] Chang C C, Chen H H, Chang Y C, et al. Computer-aided diagnosis of liver tumors on computed tomography images[J]. Computer methods and programs in biomedicine, 2017, 145: 45-51.

[10] Wei Y, Yang M, Zhang M, et al. Focal liver lesion diagnosis with deep learning and multistage CT imaging[J]. Nat Commun, 2024,15(1):7040.

[11] Solomon G B, Abebe A G, Teferi D S. Detection and classification of liver cancers using computed tomography images[Z]. Research Square, 2022.DOI:10.21203/rs.3.rs-2242020/v1

[12] Yasaka K, Akai H, Abe O, et al. Deep Learning with Convolutional Neural Network for Differentiationof Liver Masses at Dynamic Contrast-enhanced CT: A PreliminaryStudy[J]. Radiology, 2017,286(3):887-896.

[13] Liu SY. Development Report on Artificial Intelligence in Chinese Medical Imaging (2021-2022) [M]. Beijing: People's Medical Publishing House, 2022. (in Chinese)

[14] Wang L, Zhang L, Jiang B, et al. Clinical application of deep learning and radiomics in hepatic disease imaging: a systematic scoping review[J]. Br J Radiol, 2022,95(1136):20211136.

[15] Wankar B R, Kshirsagar N V, Jadhav A V, et al. Innovative Deep Learning Approach for Parkinson's Disease Prediction: Leveraging Convolutional Neural Networks for Early Detection[J]. EAI Endorsed Transactions on Pervasive Health and Technology, 2024,10.DOI:10.4108/eetpht.10.6190

[16] Krishan A, Mittal D. Ensembled liver cancer detection and classification using CT images[J]. Proc Inst Mech Eng H, 2021,235(2):232-244.

[17] Phan D, Chan C, Li A A, et al. Liver cancer prediction in a viral hepatitis cohort: A deep learning approach[J]. International Journal of Cancer, 2020,147(10):2871-2878.

[18] K. M N, A. M, S. U. Automatic Liver Cancer Detection Using Deep Convolution Neural Network[J]. IEEE Access, 2023,11:94852-94862.

[19] Manjunath R V, Ghanshala A, Kwadiki K. Deep learning algorithm performance evaluation in detection and classification of liver disease using CT images[J]. Multimed Tools Appl, 2023:1-18.

[20] Yang Y H, Zhou Z G, Li Y. MRI-based Deep Learning Model for Differentiation of Hepatic Hemangioma and Hepatoblastoma in Early Infancy[J]. European Journal of Pediatrics, 2023, 182:4365-4368.

[21] Zhang R, Lu W, Wei X, et al. A progressive generative adversarial method for structurally inadequate medical image data augmentation[J]. IEEE Journal of Biomedical and Health Informatics, 2021, 26(1): 7-16.

[22] Pu Y, Han Y, Wang Y, et al. Fine-grained recognition with learnable semantic data augmentation[J]. IEEE Transactions on Image Processing, 2024, 33: 3130-3144.

[23] Yoo J, Ahn N, Sohn K A. Rethinking data augmentation for image super-resolution: A comprehensive analysis and a new strategy[C]//Proceedings of the IEEE/CVF conference on computer vision and pattern recognition. 2020: 8375-8384.

[24] Lee H, Lee H, Hong H, et al. Classification of focal liver lesions in CT images using convolutional neural networks with lesion information augmented patches and synthetic data augmentation[J]. Medical physics, 2021, 48(9): 5029-5046.

[25] Y. Zhou, D. Dreizin, Y. Wang, F. Liu, W. Shen and A. L. Yuille, "External Attention Assisted Multi-Phase Splenic Vascular Injury Segmentation With Limited Data," in IEEE Transactions on Medical Imaging, vol. 41, no. 6, pp. 1346-1357, June 2022, doi: 10.1109/TMI.2021.3139637.

[26] Mulé S, Lawrance L, Belkouchi Y, et al. Generative adversarial networks (GAN)-based data augmentation of rare liver cancers: The SFR 2021 Artificial Intelligence Data Challenge[J]. Diagnostic and interventional imaging, 2023, 104(1): 43-48.

[27] Cristóbal I, Sanz-Álvarez M, Luque M, et al. The Role of MicroRNAs in Hepatoblastoma Tumors[J]. Cancers, 2019,11(3):409.

[28] Waters A M, Mathis M S, Beierle E A, et al. A Synopsis of Pediatric Patients With Hepatoblastoma and Wilms Tumor: NSQIP-P 2012-2016[J]. Journal of Surgical Research, 2019,244:338-342.

[29] Yasaka K, Akai H, Abe O, et al. Deep Learning with Convolutional Neural Network for Differentiation of Liver Masses at Dynamic Contrast-enhanced CT: A Preliminary Study[J]. Radiology, 2017,286(3):887-896.

[30] Vieira S, Pinaya W H L, Mechelli A. Using deep learning to investigate the neuroimaging correlates of psychiatric and neurological disorders: Methods and applications[J]. Neurosci Biobehav Rev, 2017,74(Pt A):58-75.

[31] Lucas B, Ravishankar S, Pateva I. Pediatric Primary Hepatic Tumors: Diagnostic Consider-ations[J]. Diagnostics, 2021,11(2):333.

[32] Pan EY, et al. Pediatric Imaging Diagnosis [M]. Beijing: People's Medical Publishing House, 2007. (in Chinese)

[33] LeGout J D, Bolan C W, Bowman A W, et al. Focal Nodular Hyperplasia and Focal Nodular Hyperplasia–likeLesions[J]. RadioGraphics, 2022,42(4):1043-1061.

[34] Denk H, Pabst D, Abuja P M, et al. Senescence markers in focal nodular hyperplasia of the liver: pathogenicconsiderations on the basis of immunohistochemical results[J]. Mod Pathol, 2022,35(1):87-95.

[35] Nguyen B N, Flejou J F, Terris B, et al. Focal nodular hyperplasia of the liver: a comprehensive pathologic study of 305 lesions and recognition of new histologic forms[J]. Am J Surg Pathol, 1999,23(12):1441-1454.

[36] O'Neill A F, Meyers R L, Katzenstein H M, et al. Children's Oncology Group's 2023 blueprint for research:

Liver tumors[J]. Pediatr Blood Cancer, 2023,70 Suppl 6(Suppl 6):e30576.

[37] McCarville M B, Roebuck D J. Diagnosis and staging of hepatoblastoma: Imaging aspects[J]. Pediatric Blood & Cancer, 2012,59(5):793-799.

[38] Wang XX, Zhong YM, Yuan XY, et al. CT findings of mesenchymal hamartoma of the liver in children [J]. Chinese Journal of Medical Imaging Technology, 2017, 33(9): 1288-1292. (in Chinese)

[39] Shao JB, et al. Chinese Imaging Differential Diagnosis - Pediatric Volume [M]. Beijing: People's Medical Publishing House, 2024. (in Chinese)

[40] Lee DY, et al. Pediatric Imaging Diagnosis [M]. Beijing: China Science and Technology Press, 2021. (in Chinese)

[41] Galdran A, Carneiro G, González Ballester M A. Balanced-mixup for highly imbalanced medical image classification[C]//International Conference on Medical Image Computing and Computer-Assisted Intervention. Cham: Springer International Publishing, 2021: 323-333.

[42] Jocher, G. et al. (2023). YOLOv8: Ultralytics Open-Source Object Detection Architecture. https://github.com/ultralytics/ultralytics

## Acknowledgements

This study's financial support was derived from various sources, including the Young Scientists Fund of the National Natural Science Foundation of China (82402376), the National Key Research and Development Program of China (2024YFC2419300), the Knowledge Innovation Specialized Project of Wuhan Science and Technology Bureau (2023020201010198), the Basic and Applied Basic Research Foundation of Guangdong Province (2024A1515011750), Wuhan Natural Science Foundation (2025040601020205), Hubei Provincial Association of Preventive Medicine (2025SWGKY214), and the Wuhan Children's Hospital Doctoral Start-up Fund Project (2024FEBSJJ005).

## Author Contributions

W.W., C.Y., and L.H. conceived and designed the study. W.W, C.Y. and J.S. did the investigation and acquired the data. C.Y. and W.W. did the statistical analyses. W.W., L.H. and Y.Z. developed, trained and applied the artificial neural network. C.Y., J.S., X.P., C.X, and L.H. implemented quality control of data and the algorithms. C.Y. and J.S. accessed and verified the underlying raw data. W.W. and C.Y. prepared the first draft of the manuscript. L.H., L.L. and J.S. did the project administration and supervised and coordinated the work of the research team and provided senior oversight. J.S. provided server support. All authors read and approved the final version of the manuscript. All authors were responsible for the final decision to submit the manuscript for publication.

## Competing Interests

The authors declare no competing interests.

## Appendix Table

| No. | Content |
|---|---|
| 1 | Centripetal enhancement. |
| 2 | Central stellate scar. |
| 3 | Mass density is basically consistent with the liver in the hepatic parenchymal phase and venous phase, and the mass density is roughly equal to that of the liver parenchyma in the delayed phase. |
| 4 | Cystic multilocular septated mass. |
| 5 | No obvious enhancement on contrast-enhanced scans. |
| 6 | Septal enhancement visible within the mass. |
| 7 | Obvious heterogeneous enhancement in the early arterial phase, contrast agent washout visible in the delayed phase, with density slightly lower than that of normal liver parenchyma. |
| 8 | Slight enhancement on contrast-enhanced scans. |
| 9 | No obvious enhancement in the early phase, centripetal enhancement on delayed scans. |
| 10 | Obvious enhancement on contrast-enhanced scans. |
| 11 | Nodular enhancement visible within the mass, with no obvious peripheral enhancement. |
| 12 | Peripheral enhancement in the early phase, centripetal enhancement on delayed scans. |
| 13 | Peripheral enhancement in the early phase, centripetal enhancement on delayed scan, with non-enhancing areas visible inside. |
| 14 | Peripheral enhancement in the early phase, no obvious enhancement on delayed scans. |
| 15 | Progressive enhancement. |